\documentclass{article}
\pdfoutput=1
\usepackage{trcover_unito}
\usepackage{graphicx,xspace,url}
\usepackage{latexsym,amsfonts,amssymb,pifont,booktabs,listings,xspace,color}
\usepackage{listings}
\usepackage[leqno]{amsmath}
\newcommand {\ff}{FastFlow\xspace}

\newcommand{\helv}[1]{\textsf{\small #1}\xspace}

\newcommand{\AT}{\mathcal{A}}
\newcommand{\LT}{\mathcal{L}}

\newcommand{\LeftPat}{P}

\newcommand{\RightPat}{O}

\newcommand{\srewrites}[1]{\stackrel{#1}{\longmapsto}}

\newcommand{\into}{\ensuremath{\,\rfloor}\,}

\newcommand{\conc}{\;\,}

\newcommand{\mapstoDesug}{\mapsto}

\newcommand{\red}{\mapstoDesug}

\newcommand{\ov}[1]{\overline{#1}}

\lstnewenvironment{Bench}[2]
{\lstset{ %
language=C++,                % choose the language of the code
basicstyle=\small,      % the size of the fonts that are used for the code
numbers=none,                   % where to put the line-numbers
numberstyle=\tiny,      % the size of the fonts that are used for the line-numbers
stepnumber=1,                   % the step between two line-numbers. If it's 1 each line will be numbered
numbersep=3pt,                  % how far the line-numbers are from the code
showspaces=false,               % show spaces adding particular underscores
showstringspaces=false,         % underline spaces within strings
showtabs=false,                 % show tabs within strings adding particular underscores
tabsize=2,                      % sets default tabsize to 2 spaces
captionpos=b,                   % sets the caption-position to bottom
breaklines=true,                % sets automatic line breaking
breakatwhitespace=false,        % sets if automatic breaks should only happen at whitespace
escapeinside={\%*}{*)},          % if you want to add a comment within your code
mathescape=true,
columns=flexible,
morekeywords={main,add_workers,run_then_freeze, offload,offload_eos,load_result,wait_freezing,foreach,in,from,Gillespie,stochastic_choice,find,add,put,at,to,match,against,on},
#1,
label={code:#2}
}}{}

\trnum{131/2010}

\begin{document}
%\title{Efficient  Simulations of Biological Systems on Multiprocessors}
\title{On Designing Multicore-aware Simulators for Biological Systems}

\author{Marco Aldinucci \and Mario Coppo \and Ferruccio Damiani \and 
Maurizio Drocco  \and Massimo Torquati\thanks{Dipartimento di
  Informatica, Universit\`a di Pisa, Italy} \and Angelo Troina}

\maketitle

\begin{abstract}
The stochastic simulation of biological systems is an increasingly
popular technique in bioinformatics. It often is an enlightening
technique, which may however result in being computational
expensive. We discuss the main opportunities to speed it up on
multi-core platforms, which pose new challenges for parallelisation
techniques. These opportunities are developed in two general families of
solutions involving both the single simulation and a bulk of independent
simulations (either replicas of derived from parameter
sweep). Proposed solutions are tested on the parallelisation of the
CWC simulator (Calculus of Wrapped Compartments) that is carried out
according to proposed solutions by way of the \ff programming
framework making possible fast development and efficient execution
on multi-cores.
\end{abstract}

\section{Introduction}
\label{sec:intro}

Stochastic simulations are an increasingly popular technique to study
biological systems. They -- differently from other modelling
approaches such as differential equations (ODEs) -- are able to
describe transient, and multi-stable behaviours of the systems.

Different formalisms, based on automata
models (like ~\cite{ABI01}), process algebras
(like~\cite{PRSS01,Car05}) or rewrite systems (like~\cite{P02}) have
either been applied to, or inspired from, biological
systems. %BMMTT08
Quantitative simulations of biological models represented with these kinds of frameworks (see, e.g.~\cite{PRSS01,KMT08,DPR08})
are usually developed via a stochastic method derived by Gillespie's
algorithm~\cite{G77}. %BMMTT08

Among other formalisms, the Calculus of Wrapped Compartments (CWC)
\cite{preQAPL2010} is a recently proposed rewriting-based language for
the representation and simulation of biological systems. It has been
designed with the aim of simplifying the development of efficient
implementations, while keeping the same expressiveness of other more
complex languages.  %(a system directly derived from CSC \cite{BMMTT08}) 

Stochastic simulations are computationally more  expensive than ODEs
numerical solution. This is particularly true for the kind of systems
that are better represented by stochastic models since, for their
uneven nature, should be simulated at a very fine grain to
spot possible spikes of the modelled phenomena along time, or to
discriminate families of possible behaviour that are not revealed by
the averaged behaviour described by ODEs.

The high computational cost of stochastic simulation is well known and
has led, in the last two decades, to a number of attempts to
accelerate them up using several kind of techniques, such as
approximate simulation algorithms and parallel computing
\cite{balbo:sim:1998}.  In this work, this latter approach is taken
into account.

Since stochastic simulations are basically Monte Carlo processes,
many independent instances should be computed to achieve a
statistically valid solution. These independent instances have been
traditionally exploited in an \emph{embarrassingly parallel} fashion,
executing a partition of the instances in a different machine. This
approach naturally couples with distributed computing (i.e. cluster,
grid, clouds).

However, the entire hardware industry has been moving to multi-core, which
nowadays equips the large majority of computing platforms. The rapid shift
towards multi-core technology has many drivers that are likely to
sustain this trend for several years to come. These platforms, which
are increasingly diffused in scientific laboratories, typically offer
moderate to high peak computational power. This potential power,
however,  cannot always be turned into actual application
speed. This is particularly true for I/O- and memory-bound
applications since all the cores usually share the same memory and
I/O subsystem.

The analysis of biological systems produces a large amount of data,
often organised in streams coming from either analysis instruments or
simulators. The management of these streams in not
trivial on multi-core platforms as the memory bandwidth cannot usually
sustain a continuous flux of data coming form all the cores at the
same time.

A related aspect regards analysis of the simulation results, which
requires the merging of results from different simulation instances and possibly
their statistical filtering or mining. In distributed computing, this
phase is often demoted  to a secondary aspect in the computation and
treated as with off-line post-processing tools. However, this approach
is no longer realistic because of both 1) the ever-increasing size of
produced data and, 2) it insists on the main weakness of
multi-core platforms, i.e. memory bandwidth and core synchronisations.

In this paper we propose a critical rethinking of the parallelisation
of stochastic processes in the light of emerging multi-core platforms
and the tools that are required to derive an efficient simulator from
both performance and easy engineering viewpoints. We believe that this latter aspect
is of crucial importance for next generation biological tools because
they will be largely designed by bioinformatic scientists, who will be
certainly much more interested in the accurate modelling of natural
phenomena rather than on the synchronisation protocols required to build
efficient tools on multi-core platforms.

We use the CWC calculus and its sequential
simulator  (Sec. ~\ref{sec:cwc})  as paradigmatic example
to discuss the key features required to derive an easy porting on a
multi-core platform (Sec.~\ref{sec:par}). In particular we will
argument on the parallelisation of a single simulation instance
(Sec.~\ref{sec:speeding_single}), many independent instances
(Sec.~\ref{sec:manysims}), and the technical challenges they
require. Among these, parallel programming tools and frameworks for multi-core are
discussed in Sec.~\ref{sec:stream}, in particular we will focus on
stream oriented pattern-based parallel programming supported by the
\ff framework (Sec.~\ref{sec:fastflow}).

The key features discussed in Sec. ~\ref{sec:cwc} are turned into a
family of solutions to speed up both the single simulation instance
and many independent instances. The former issue is approached using
SIMD hardware accelerators (Sec.~\ref{sec:implsingle}), the latter
advocating a novel simulation schema based on \ff accelerator that
guarantees both easy development and efficient execution
(Sec.~\ref{sec:implmany}). The proposed solutions are experimentally
evaluated.

\section{The Calculus of Wrapped Compartments}
\label{sec:cwc}

The Calculus of labelled Wrapped Compartments (CWC) (a small variant of the one presented in \cite{preQAPL2010}) is based on a nested structure of ambients delimited by membranes with specific proprieties. Biological entities like cells, bacteria and their interactions can be easily described in CWC.

\subsection{CWC}
\label{CWC_formalism}

Let $\AT$ be a set of  \emph{atomic elements} (\emph{atoms} for
short), ranged over by $a$, $b$, ..., and  $\LT$ a set of \emph{compartment types} represented as \emph{labels} ranged over by $\ell,\ldots$.

A \emph{term} of CWC is a multiset $\ov{t}$ of  \emph{simple terms} where a simple term  is either an atom $a$ or a compartment $(\overline{a}\into
\overline{t'})^\ell$ consisting of a \emph{wrap} (represented by the multiset of atoms $\overline{a}$), a \emph{content} (represented by the term
$\overline{t'}$) and a \emph{type} (represented by the label $\ell$).

An example of term is $a \conc b \conc (c \conc d \into e \conc f)^\ell$ representing a multiset (multisets are denoted by listing the elements separated by a space) consisting of two atoms $a$ and $b$ (for instance two
molecules) and an $\ell$-type compartment $(c \conc d \into e \conc f)^\ell$ which, in turn, consists of a wrap (a membrane) with two atoms $c$ and $d$
(for instance, two proteins) on its surface, and containing the atoms $e$ (for instance, a molecule) and $f$ (for instance a DNA strand).

System transformations are defined by rewriting rules. A rewriting rule is defined as a pair of terms (on an extended set of atomic elements which includes variables), which represent the \emph{patterns}, ranged over by $\LeftPat,\; \RightPat$, together
with a label $\ell$ representing the compartment type to which the rule can be applied. Rules are represented as expression of the form $\ell:\LeftPat \red \RightPat$.
A simple example of a rewrite rule is
 $$\ell:~a \; b \; X \red c\; X $$
meaning that in all compartments of type $\ell$ an occurrence of $a,\, b$ ($X$ can match with all the remaining part of the compartment content) can be replaced by $c$.

The application of a rule $\ell:\LeftPat \red \RightPat$ to a term
$\ov{t}$ is performed in the following way:\\
\begin{itemize}
 \item find (if it exists) the content (or the wrap) $\ov{u}$ of a compartment of type $\ell$ in $\ov{t}$ and an substitution $\sigma$ of variables by terms such that $\ov{u} = \sigma(\LeftPat) $.
 \item  replace in $\ov{t}$ the subterm $\ov{u}$ with
   $\sigma(\RightPat)$.
\end{itemize}
We write  $\ov{t} \red \ov{t'}$ if $\ov{t'}$ is obtained  by applying a rewrite rule to  $\ov{t}$.

\subsection{Stochastic Simulation}\label{SECT:STO_SEM}

A stochastic simulation model for biological systems can be defined by incorporating a collision-based stochastic framework along the line of the one
presented by Gillespie in \cite{G77}, which is, \emph{de facto}, the standard way to model quantitative aspects of biological systems. The idea of
Gillespie's algorithm is that a rate constant is associated with each considered chemical reaction. Such a constant is obtained by multiplying the kinetic
constant of the reaction by the number of possible combinations of reactants that may occur in the system (thus modelling the law of mass action, but more flexible approaches are also considered in the literature \cite{preQAPL2010}). The resulting rate is then used as the parameter of an exponential distribution modelling the time spent between two occurrences of the considered chemical reaction.

Each reduction rule is enriched by the kinetic constant $k$ of the reaction that it represents (notation $\ell:\LeftPat \srewrites{k} \RightPat $). The number of reactants in a reaction represented by a rewrite rule is evaluated considering the number of distinct occurrences, in the same context, of subterms matching with the considered rule.
For instance in evaluating the application rate of the stochastic rewrite rule $R=
\ell: a \conc b \conc X \srewrites{k} c \conc X$ to the term $\ov{t}=a\conc a \conc b \conc b$ in a compartment of type $\ell$ we must consider the number of the possible combinations of reactants of
the form $a\conc b$ in $\ov{t}$. Since each occurrence of $a$ can react with each occurrence of $b$, this number is 4. So the application rate of $R$ is
$k\cdot 4$.
This number can be evaluated by specific algorithms (we refer
to\cite{preQAPL2010} for a more detailed account).  %\cite{BMMTT08},
The stochastic simulation algorithm is
essentially a \emph{Continuous Time Markov Chain} (CTMC).
Given a term $\ov{t}$, a set ${\cal R}$ of reduction rules, a global time $\delta$ and all the reductions $e_1,\ldots,e_M$ applicable to $\ov{t}$, with rates $r_1,\ldots,r_M$ such that $r=\sum_{i=1}^M
r_i$, the standard simulation
procedure that corresponds to Gillespie's simulation algorithm~\cite{G77} consists of the following two steps:

\begin{enumerate}
\item The time $\delta+\tau$ at which the next stochastic reduction will occur is randomly
chosen with $\tau$ exponentially distributed with parameter $r$;
\item The reduction $e_i$ that will occur at time $\delta+\tau$ is randomly
chosen with probability ${r_i}/{r}$.
\end{enumerate}

\subsection{The CWC simulator}
\label{sec:simulator}
% By definition,
% Gillespie algorithms iterates 1) a \emph{Monte Carlo step}, where  the
% next reaction to occur is (somehow) selected according to a random
% number; 2)  an \emph{update step} increase the time step by the
% randomly generated time in step 1, and (somehow) update the molecule
% count based on the reaction that occurred. 

The CWC simulator is a tool strictly based on Gillespie
algorithm \cite{cwc:web}. It iterates the following three logical steps:
\begin{enumerate}
\item
  \emph{Match}: it searches for the
  occurrences of the rules (object
  trees) inside the term (subject tree),
  according to the notion of contexts.
  Then it associates a stochastic rate
  to each match. This step results into
  a weighted matchset.
\item
  \emph{Resolve (Monte Carlo step)}: it stochastically
  decides the \emph{time} (offset)
  at which the next reaction will occur
  and the \emph{rule} that
  will activate it.
  Moreover, since in CWC reactions can
  occur at different contexts, it
  consults the matchset in order to
  decide how portion of the system will react.
\item
  \emph{Update}: if effectively
  applied the selected reaction,
  affecting both the system and the
  clock, moving forward the simulation process.
\end{enumerate}

\section{Exploiting Parallelism in Stochastic Simulations}
\label{sec:par}
\begin{figure}
\begin{center}
\includegraphics[width=0.75\linewidth]{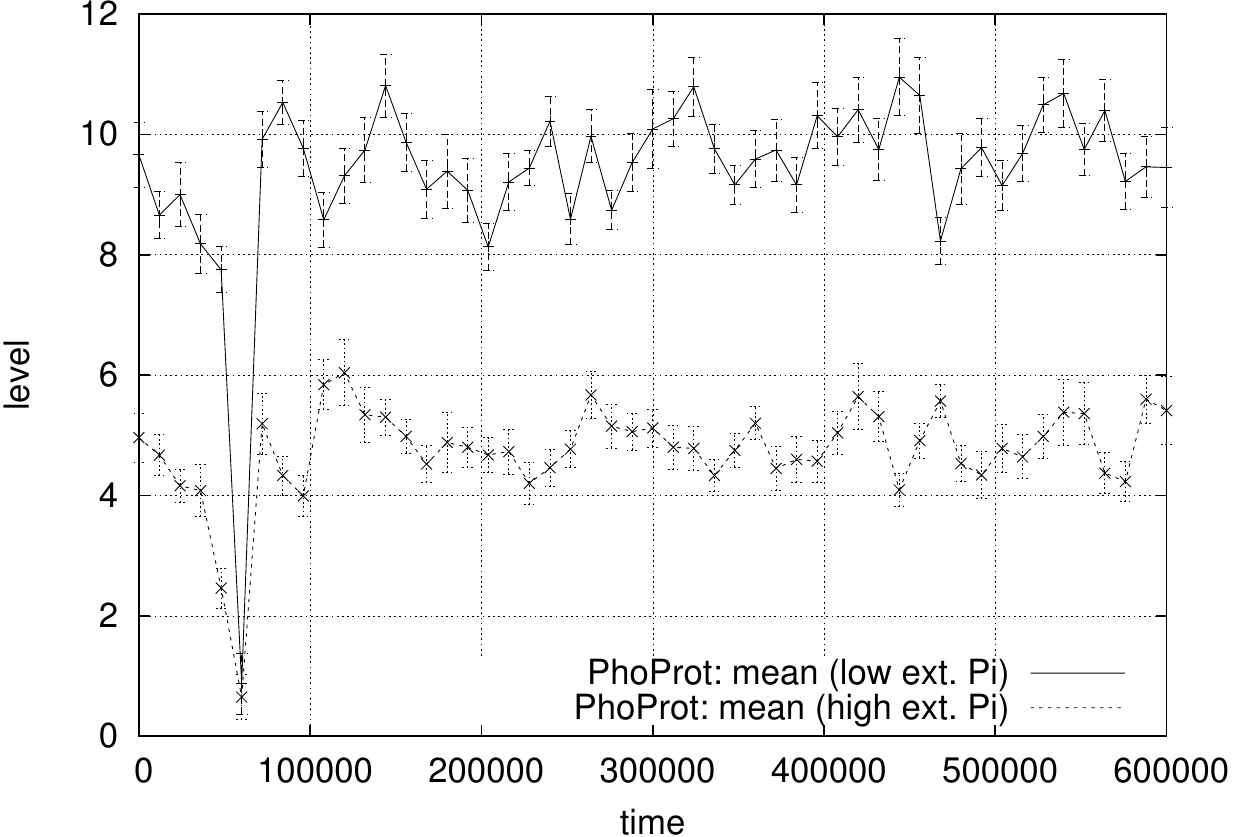}
\caption{Output of the CWC simulator for gene regulation in E. Coli
  model: average of 100 independent instances with variance (90\%
  confidence) computed a fixed simulation time steps.\label{fig:simout}}
\end{center}
\end{figure}

% Stochastic simulation algorithms and methods were initially
% developed to analyse chemical reactions involving large numbers of
% species with complex reaction kinetics. The first algorithm of this
% kind, which was proposed by Gillespie in 1977, is an exact procedure
% for numerically simulating the time evolution of a chemically reacting
% system \cite{G77}. 
Gillespie algorithm realises a  Monte Carlo type
simulation method, thus it relies on repeated random sampling to
compute the result. 
% As an example, in reaction kinetics the idea is
% to start with some amount of each species at time zero, then advance
% through time in small random steps, using a random number to decide
% which reaction should occur at each time-step until the simulated time
% reaches a given stopping time. 
An individual simulation, which tracks
the state of the system at each time-step, is called a
\emph{trajectory}. Each individual trajectory represents just one possible
way in which the system might have reacted over the time-span from the
start-time to the stop-time. Many thousands of trajectories
might be needed to get a representative picture of how the system behaves on the
whole. An example of the synthesis of many trajectories obtained from
the CWC simulator is reported in  Fig.~\ref{fig:simout}. In the
figure, the two curves are obtained by averaging 100 trajectories from independent simulation instances  at fixed simulation
time steps and computing variance with 90\% confidence intervals over them.

% The stochastic simulation  is an informative technique to study biological
% system since -- differently from other modelling approaches such as differential
% equations (ODEs) -- are able to describe transient, unstable and multi-stable
% behaviours. 

For this, stochastic simulations are computationally more 
expensive than ODEs numerical unfolding. This balance is well-known
and it motivated many attempts to speed up their
execution time along last two decades \cite{balbo:sim:1998}. 
They can be roughly categorised in attempts that
tackle the speeding up of a \emph{single simulation} and a \emph{bulk
  of independent simulations}. In the following these (not mutually exclusive)
approaches are discussed under the viewpoint of parallel computing
techniques and their exploitation on commodity multi-core
platforms. This discussion is not intended to be an encyclopaedic
review of other techniques that can be used to achieve the same aim,
such as ones related to the approximation of the simulation results,
such as $\tau$-leaping and  hybrid techniques \cite{ABI01}.  

\subsection{What can speeded up? Where parallelism can be found?}
\subsubsection{Speeding up a  single simulation} 
\label{sec:speeding_single}
Parallelising a single Gillespie-like stochastic simulation, i.e. the derivation of 
a simulation trajectory,  is intrinsically hard. Unless introducing
algorithmic relaxations -- which correctness should be proved and
typically lead to algorithm approximations -- two successive Monte
Carlo steps cannot be concurrently executed since there exists a strict
data dependency between the two steps. Also, at the single step grain,
speculative execution is 
unfeasible because of the excessive branching of possible future
execution paths.\footnote{Even if also this kind of approach has been
  attempted and it might result feasible whether coupled with algorithm relaxations \cite{balbo:sim:1998}.} As result, the only viable option to exploit parallel
computing within a single simulation consists in parallelising the
single Monte Carlo step. Here, the available concurrency could be
determined via data dependency analysis
\cite{bernstein:66} that can be made for any given
specific simulator code (see Sec.~\ref{sec:simd:cwc}). Typically,
parallelism exploited at this level is extremely fine-grained since the
longest concurrent execution path may at most  count dozens to
hundreds machine instructions. 
% for an expected execution time of (at most) few hundreds of nanoseconds. 

In this range, currently, no software mechanisms can support an
effective inter-core or multi-processor parallelisation: the
overhead will easily overcome any benefit; the only
viable option is hardware parallelism within a single core. Since,
typically, instruction stream parallelism is already exploited by
superscalar processor architecture, the only additional
parallelisation opportunity has to be searched in data parallelism to
be exploited via a hardware accelerator, such as internal SSE/MMX or
external GPGPU accelerators (General-Purpose GPU). In both cases, the
simulator code should be 
deeply re-designed in a contiguous sequence of SIMD
instructions.  As we shall see in Sec.~\ref{sec:simd:cwc}, this
generally may lead to very modest advantages with respect to the required effort.

\subsubsection{Speeding up independent simulation instances}
\label{sec:manysims}
The intrinsic complexity in the parallelisation of the single step has
traditionally led to the exploitation of parallelism in the
computation of independent instances of the same simulation, which
should anyway be computed to achieve statistical convergence of
simulated trajectories (as in all Monte Carlo methods). The problem is
well understood; it has been exploited in the last two decades in
many different flavours and distributed computing
environments, from clusters to grid to clouds. Notwithstanding that
the problem has been often approached in a simplified form, often
assuming that output data has a negligible size, as it happens in Monte
Carlo Pi computation; this is not likely to happen in this and next generation
biological simulations.

In particular, simulation distribution, result gathering,
trajectory data assembling and analysis phases are neither considered as a
part of the problems to be accelerated nor considered in the
performance evaluation. As matter of a fact, parallel simulations  is
often considered an ``embarrassingly parallel'' problem, whereas it is
-- if and only if -- data distribution, gathering, filtering, and analysis
are not considered as part of the whole process.  
Unfortunately, it happens that they may result as expensive as the
simulation itself. As an 
example, a simulation of the HIV diffusion problem
(computed using the StochKit toolkit for 4 years of simulation time) produces about 
5 GBytes of data per instance 
\cite{stochkit-ff:tr-10-12}. As clear, the data size is
$n$-folded when $n$ instances are considered. During post-processing
phase, this data should be gathered and often reduced to a single
trajectory via statistical methods. 

These potential performance flaws are further exacerbated in multi-core and
many-core platforms. %%, which are becoming increasingly popular. 
These platforms do not exhibit the same degree of
replication of hardware resources that can be found in distributed
environments, and even independent processes actually compete for the
same hardware resources within the single platform, such main and
secondary memory, which performances represent the real challenge of
forthcoming parallel programming models (a.k.a. \emph{memory wall}
problem). While simulation is substantially a CPU-bound problem on
distributed platform, it may become prevalently an I/O-bound problem
on a multi-core platform due to the need to store and post-process
many trajectories. In particular, multi-stable simulations may require
very fine grain resolution to discriminate trajectory state changes,
and as it is clear, the finer the observed simulation
time-step the strongest the computational problem is characterised as
I/O-bound. 

%\subsection{Guidelines for the effective parallelisation on
%multi-core}
\subsection{How to parallelise? A list of guidelines for the effective
  parallelisation on multi-core}
% In Sec.~\ref{sec:speeding_single} we claimed that parallelism
% available in a \emph{single step} of Gillespie-like algorithms is
% definitely too fine-grained to be exploited among different cores,
% whereas parallelism available in \emph{independent instances}  
In the previous section we discussed where parallelism can be found in
Gillespie-like algorithms; the question that naturally follows is \emph{how}
this parallelism can be \emph{effectively} exploited.  We advocate here a number of
parallelisation issues that, we believe, can 
be used as pragmatic ``guidelines'' for the efficient parallelisation of this kind
of algorithms on multi-core.  Observe that, in principle, they are
quite independent of the source of parallelism; however, they focus on
inter-core parallelism, thus cannot be expected to be applied to other
kinds of parallelism (e.g. SIMD parallelism). They will be then used along
Sec.~\ref{sec:simd:cwc} as ``instruments'' to evaluate the quality of the
parallelisation work for the execution of independent instances of the
CWC simulator.
% fine-grained parallelism (e.g. Gillespie's single step) that simply
% cannot be exploited on different cores.

%be hardly applied to single step parallelism
% because thus  Since these issues are
% mainly related to inter-core parallelism, we expect they can be hardly effective

% The categorisation of parallelisation approaches focusing at \emph{single step}
% (Sec.~\ref{sec:speeding_single})  and \emph{independent instances}
% (Sec.~\ref{sec:manysims}) is quite established and universal (as it can
% be applied to different parallel platforms).  As we di
% but that might fall short
% in providing a pragmatic guide to the design and porting of efficient
% stochastic simulator on multi-core platforms. In this regard, we
% advocate here a number of parallelisation issues that, we believe, can
% be used as ``guidelines'' for the efficient parallelisation of this kind
% of algorithms on multi-core. They will be then used along
% Sec.~\ref{sec:simd:cwc} as ``instruments'' to evaluate the quality of the
% parallelisation work for the CWC simulator.

\subsubsection{Data stream as a first-class concept} The \emph{in
  silico} (as well as \emph{in vitro}) analysis of
biological systems produces a huge amount of data. Often, they can be
conveniently represented as data streams since they sequentially flows out
from one or more (hardware or software) devices; often the cost of
full storage of these streams overcomes their utility, as in many cases
only a statistical filtering of the data is needed. These data
streams can be conveniently
represented as \emph{first-class concept}; their management
should be performed  \emph{on-line} by exploiting the potentiality of
underlying multi-core platforms via a suitable \emph{high-level programming tools}. 

\subsubsection{Effective, high-level programming tools}  To date,
parallel programming has not 
embraced much more than low-level communication and synchronisation
libraries. In the hierarchy of abstractions, it is only slightly above
toggling absolute binary in the front panel of the machine. We
believe that, among many, one of the reasons for such failure is
the fact that programming multi-core is still perceived as a branch
of high-performance computing with the consequent excessive focus on
absolute performance measures. By definition, the raison d'{\^e}tre
for high-performance computing is high performance, but MIPS, FLOPS
and speedup need not be the only measure. Human productivity, total
cost and time to solution are equally, if not more, important. The shift to multi-core
is required to be graceful in the short term: existing applications should be ported to
multi-core systems with moderate effort. This is particularly
important when parallel computing serves as tools for other sciences
since non expert designer should be able to experiment different
algorithmic solutions fro both simulations and data analysis. This
latter point, in particular, may require data synchronisation and
could represent a very critical design point for both correctness and
performance.  

\subsubsection{Cache-friendly synchronisation for data
streams} Current commodity multi-core and many-core platforms exhibit
a cache-coherent shared memory since it makes it can effectively reduce
the programming complexity of parallel programs (whereas different
architectures, such as IBM Cell, have exhibited their major limits in
programming complexity). Cache coherency is not for free, however. It
largely affects synchronisations cost and may require expensive
performance tuning. This is both an opportunity and a challenge for
parallel programming framework designers since a properly designed
framework should support the application with easy exploitation of
parallelism (either design from scratch or porting from sequential
code) and high-performance.

\subsubsection{Load balancing of irregular workloads} Stochastic
processes exhibit an irregular behaviour in space and time by their
very nature since different simulations may cover the same simulation
timespan following many different, randomly chosen, paths and number
of iterations. Therefore, parallelisation tools should support the
dynamic and active balancing of workload across the involved cores.

\section{Pattern-based high-level stream parallelism}
\label{sec:stream}

\emph{Stream parallelism} is a programming paradigm supporting the
parallel execution
of a stream of tasks by using a series of \emph{sequential} or
\emph{parallel} stages.  A stream program can be naturally represented
as a graph of independent \emph{stages} (kernels or filters)
that communicate explicitly over data
channels.  Conceptually, a streaming computation represents a sequence
of transformations on the data streams in the program.  Each stage of
the graph reads one or more tasks from the input stream, applies some
computation, and writes one or more output tasks to the output stream.
Parallelism is achieved by running each stage of the graph
simultaneously on \emph{subsequent} or \emph{independent} data
elements.
%
%Local state may be either maintained in each stage or distributed
%(replicated or scattered) along streams.

As with all kinds of parallel program, stream programs can be expressed as a
graph of concurrent activities, and directly programmed using a
low-level shared memory or message passing programming
framework. Although this is still a common approach, writing a correct,
efficient and portable program in this way is a non-trivial
activity. 
Attempts to reduce the programming effort by raising the level
of abstraction through the provision of parallel programming frameworks date back at
least three decades and have resulted in a number of significant contributions.
Notable among these is the \emph{skeletal}
approach \cite{cole-th} (a.k.a. \emph{pattern-based}
parallel programming), which appears to be becoming increasingly popular
after being revamped by several successful parallel programming
frameworks
\cite{mapreduce:google:04,intel:skeletons:tbb}.

\emph{Skeletons} (a.k.a. patterns)
capture common parallel programming paradigms (e.g. MapReduce,
ForAll, Divide\&Conquer, etc.) and make them available to the
programmer as high-level programming constructs equipped with
well-defined functional and extra-functional semantics
\cite{lithium:sem:CLSS}.
%Some of these attempts explicitly include stream parallelism as a major source of
%concurrency exploitation
%\cite{van:assist:02,lithium:sem:CLSS,streamIt,intel:skeletons:tbb}:
%rather than allowing programmers to connect stages into arbitrary
%graphs, basic forms of stream parallelism are provided to the
%programmer in high-level constructs such as \emph{pipeline},
%\emph{farm}, and \emph{loop}.

The \emph{pipeline} skeleton is one of the most widely-known, although
sometimes it is underestimated. Parallelism is achieved by running
each stage simultaneously on subsequent data elements, with the pipeline's
throughput being limited by the throughput of the slowest stage.

The \emph{farm} skeleton models functional replication and consists in running
multiple independent stages in parallel, each operating on different tasks of
the input stream. The farm skeleton is typically used to improve the
throughput of slow stages of a pipeline.
It can be better understood as a three stage -- \emph{emitter},
 \emph{workers}, \emph{collector} -- pipeline. The emitter
dispatches stream items to a set of workers, which independently
elaborate different items. The output of the workers is then gathered by
the collector into a single stream. These logical stages are considered
by a consolidated literature as the basic building blocks of stream
programming.
%However, they admit
%many aliases and variants.
%In the \emph{master-worker} pattern the emitter and the
%collector are wrapped up in a single logical master (either centralized or distributed).
%
%Emitter, workers and collector are
%also known as \emph{split, filter, join}  \cite{streamIt};
%\emph{scatter, operate, gather} \cite{stream:micro:05}; and
%\emph{in\_sec\-tion, parmod, out\_sec\-tion}
%\cite{van:assist:02}, respectively.
%

% Also, they admit different dispatching (unicast, multicast, scatter) and
% gathering (from\_any, from\_all) policies  that can be characterized  as static/dynamic,
% in-order/out-of-order. Workers may exhibit a
% permanent local state, global state (read-only, read-write), or
% transient state.

The \emph{loop} skeleton (also known as \emph{feedback}), provides a
way to generate cycles in a stream graph. This skeleton is typically
used together with the farm skeleton to model recursive and
Divide\&Conquer computations.

%As will be seen in the next sections, the stream paradigm perfectly suits
%the need for reducing inter-core synchronization overheads in parallel
%programs for shared cache multi-cores. Therefore, it can be used to
%build an efficient run-time support for a high-level programming model
%aimed at the effective design of parallel applications.
%In this work, a pattern-based approach
%to support the acceleration of sequential codes is advocated,
%where patterns model a concurrent
%management of control and data streams, which may be available as
%application input or auto-generated within the application itself.

In particular, the \emph{\ff} implementation of the loop and farm
patterns will be exploited in Sec.~\ref{sec:simd:cwc} to parallelise
the CWC simulator.  

\subsection{The \ff skeleton-based programming framework}
\label{sec:fastflow}
\ff is a C++ parallel programming framework aimed at {simplifying} the
development of {efficient} applications for multi-core platforms.
The key vision of \ff is that
ease-of-development and runtime efficiency can both be achieved by raising
the abstraction level of the design phase, thus providing
developers with a suitable set of parallel programming patterns that
can be efficiently compiled onto the target platforms
\cite{fastflow:web}.

\begin{figure}[t!]
\begin{center}
\includegraphics[width=0.85\linewidth]{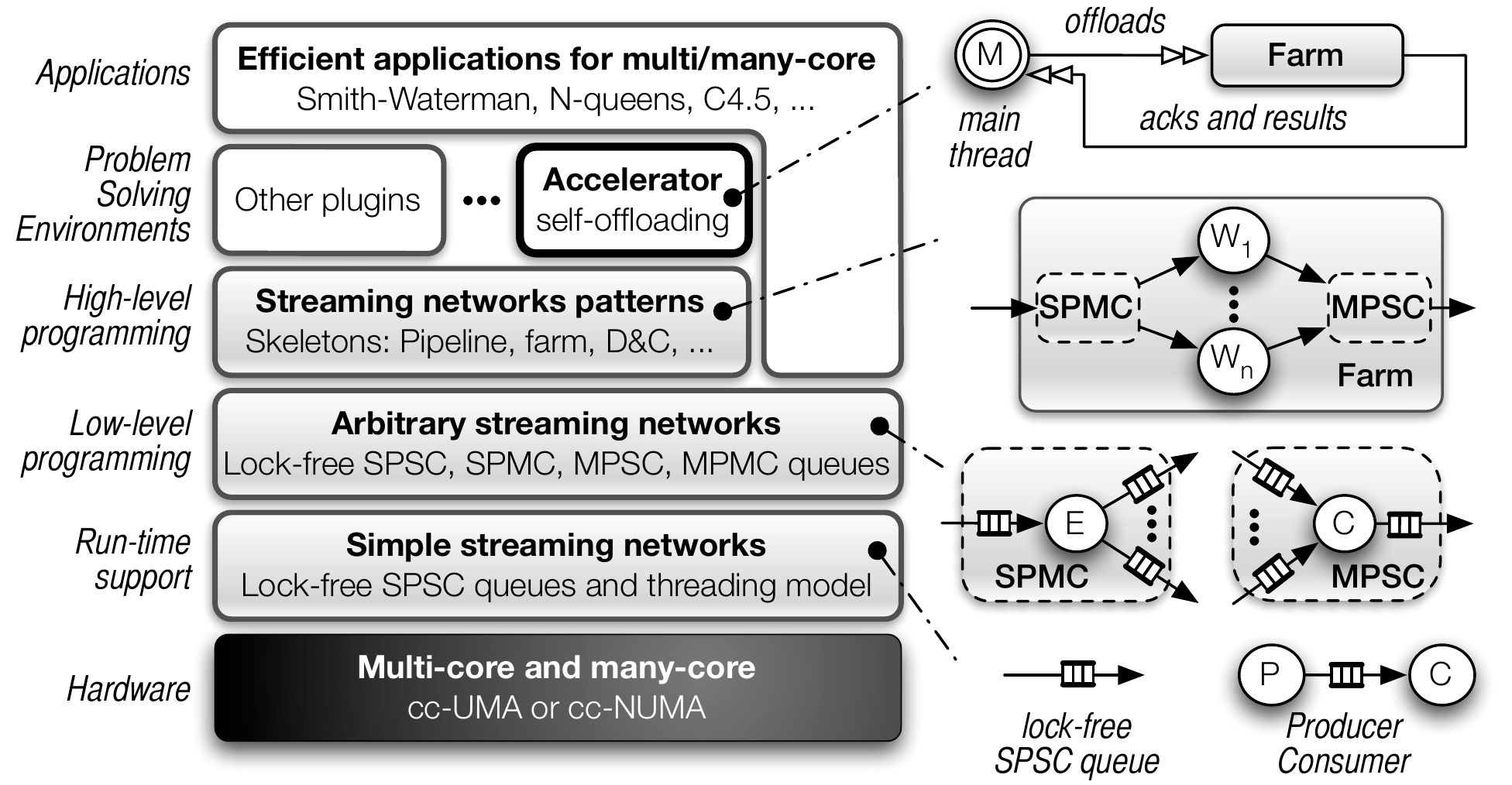}
\caption{FastFlow layered architecture with abstraction examples at
  the different layers of the stack.\label{fig:ff:architecture}}
\end{center}
\end{figure}

\ff is conceptually
designed as a stack of layers that progressively abstract the shared
memory parallelism at the level of cores up to the definition of
useful programming constructs supporting structured parallel
programming on cache-coherent shared memory multi- and many-core
architectures (see Fig.~\ref{fig:ff:architecture}).
These architectures
include commodity, homogeneous, multi-core systems
such as Intel core, AMD K10, etc.  \ff natively supports stream
parallelism since it implements parallelism patterns as data-flow graphs --
so-called \emph{streaming networks}.

The core of the \ff framework (i.e. \emph{run-time support} tier)
provides an efficient implementation of
Single-Producer-Single-Consumer
(SPSC) FIFO queues. \ff SPSC queues are lock-free, wait-free, and
do not use interlocked operations \cite{fastflow:web,Herlihy91}.
%\cite{fastforward:ppopp:08}. 
% Also, they
% do not make use of
% any memory barrier for Total Store Order processors (e.g. Intel core)
% and use a single memory write barrier (in the push operation) for
% processors supporting weaker memory consistency models
% \cite{fastflow:web}.

The SPSC queue is primarily used as  synchronisation mechanism for
memory pointers in a consumer-producer fashion. The
next tier up extends one-to-one queues (SPSC) to
one-to-many (SPMC), many-to-one (MPSC), and many-to-many (MPMC)
synchronisations and data
flows, which  are implemented using only SPSC queues and
arbiter threads, thus providing lock-free and wait-free arbitrary
data-flow graphs (\emph{arbitrary streaming networks}) that requires few
or no memory barriers, and thus few cache invalidations. 
% Also, cyclic
% graphs can be defined in such a way that they are provably deadlock-free
% \cite{fastflow:web}.

The upper layer, i.e. \emph{high-level programming},  provides a
programming framework based on parallel patterns.
% (see Sec.~\ref{sec:stream} and Sec.~\ref{sec:methodology}).
In particular, \ff  provides \emph{farm}, \emph{farm-with-feedback}
(i.e. Divide\&Conquer) and \emph{pipeline} patterns, and supports their
arbitrary nesting and composition.  The \ff pattern set can be
further extended by building new C++ templates.

\ff is available as an open source software under LGPLv3
\cite{fastflow:web}. A performance comparison against other
programming tools such as POSIX, Cilk, OpenMP, and Intel TBB has been
reported in \cite{fastflow:web,fastflow:pdp:10}.

\section{The CWC Simulator Testbed}
\label{sec:simd:cwc}

The proposed guidelines are validated using the CWC simulator as
running example. It has been developed as a plain C++ sequential code
(exploiting the C++ boost library), then it has been parallelised for
multi-core. In order to evaluate the effectiveness of the methodology
also in term of development effort,  the original code and its
parallel versions have been developed by two different (master
student) teams. In the parallelisation two main frameworks have been
used: the GCC compiler SSE intrinsics \cite{intel:sse:intrics} to
speed up a single simulation, and the \ff parallel programming
framework \cite{fastflow:web} to speed up independent simulation
instances, which provides the basic facilities described in
Sec.~\ref{sec:manysims} and that is briefly recapped in
Sec.~\ref{sec:fastflow}.

%\subsection{Experimentation platform}
All reported experiments  have been
executed on an Intel  workstation with 2 quad-core Xeon E5520
Nehalem (16 HyperThreads) @2.26GHz with 8MB L3 cache
and 24 GBytes of main memory with Linux x86\_64.
The Nehalem processor uses Simultaneous MultiThreading (SMT,
a.k.a. HyperThreading) with 2 contexts per core and the Quickpath
interconnect equipped with a distributed cache coherency protocol. SMT
technology makes a single physical processor appear as two logical
processors for the
operating system, but all execution resources are shared between the
two contexts. Each core is equipped with a SSE4.2 SIMD engine.
%
%With the exception of very long runs,
%All presented experimental results are taken as an average of 5 runs exhibiting
%low variance.

% \subsection{Simulator validation}
% %\subsection{Statistical convergence}
% Simulations processes have always to achieve some statistical validity.
% A typical example is the estimation of the mean for a certain quantity
% (e.g. the concentration of a molecule inside a cell), by expanding
% the sample mean with a \emph{confidence interval}.
% At some fixed confidence level (e.g. 90\%), the only way to make more precise
% estimations (obtaining more thin intervals) is by 
% increasing the number of simulation instances.
% For this experiment, we used a model of phosphate regulation in Escherichia Coli.

\subsection{Speeding up a  single simulation}
\label{sec:implsingle}

\begin{figure}
\begin{minipage}{\linewidth}
 \begin{Bench}{}{pseudocode}
Simulation_Step {
  // 1. Match
    foreach r $\in$ ruleset {
      Match(r, T, TOP_LEVEL); // [non-SIMD parallelism]
  // 2. Resolve (Monte Carlo)
    (tau, mu) = Gillespie(matchset);
    context = stochastic_choice on matchset[mu];
  // 3. Update
    (P,O) = left_and_right_side(mu);
    delete P_sigma from T at context; // SIMD
    put deleted_elements in sigma;
    add O_sigma to T at context; // SIMD
    simclock += tau;
}

Match(rule, term, context) {
  P = left_hand_side(rule);
  count_population =  Match_Populations(atoms(P), atoms(term)); // SIMD
  count_compartments =  match compartments(term) against compartments(P);
  stoch_rate = count_population * count_compartments * kinetic(rule);
  put (context, stoch_rate) in matchset[rule];
  foreach c_t $\in$ compartments(term)
    Match(rule, content(c_t), c_t); // recursive step [non-SIMD parallelism]
}

Match_Populations(atoms_object, atoms_subject) {
  count = 1;
  foreach (x, k) $\in$ atoms_object {
    find (x, n) in atoms_subject;
    count *= binomial(n, k);
  }
  return count;
}
\end{Bench}
\end{minipage}
\caption{CWC simulator  pseudo-code (see also Sec.~\ref{sec:simulator}) with possible sources of fine-grain
  parallelism.\label{fig:simd}}
\end{figure}

As discussed in Sec.~\ref{sec:speeding_single}, the parallelisation
of the single CWC simulation step is theoretically feasible via the SSE
accelerator. The pseudo-code of the simulation step is sketched in
Fig.~\ref{fig:simd}.
In the figure, the phases of the code that can be parallelised in SIMD
fashion with moderate effort are marked with the \emph{``SIMD''}
label. The exploited parallelism degree is 4 since 4x32-bit operation
has been used; Fig.~\ref{fig:simd:speedup} reports the achieved
speedup on a single core for $n$-species of the \emph{Lotka-Volterra}
models (the 2-species case is the standard \emph{prey-predator}
model). Despite SSE exhibits very low overhead, the achieved speedup
is almost negligible because only a fraction of the whole 
simulation  step has been actually parallelised (Amdahl's law's
applies \cite{amdahl:law:1967}).  Similar parallelisation efforts
conducted on GPGPU accelerators, which 
exploit a much larger  potential SIMD parallelism, do not actually result in 
satisfactory results. As an example, see the parallelisation of Gillespie’s first
reaction method on NVIDIA CUDA \cite{DBLP:conf/iccms/DittamoC09}.

%Also, Fig.~\ref{fig:simd:speedup}
%reports possible sources of \emph{``non-SIMD''} parallelism, actually
%Divide\&Conquer parallelism, which cannot be directly exploited in SIMD
%style.  without simulating recursion with iteration. 

Unfortunately, the extension of the  SIMD parallelism to larger fractions of
the code may require a very high coding effort since the redesign of
the original code is required. As an example recursive patterns
(used for tree-matching, marked with \emph{``non-SIMD''} parallelism
in  Fig.~\ref{fig:simd}) are not easily manageable using SIMD
parallelism and should be differently coded before being
parallelised. Observe that these recursive kernels cannot either be
parallelised across cores because they are excessively fine-grained;
as an example the parallelisation via POSIX threads (tested with \ff and Intel
TBB) is, in our the reference platform, from  $10$ to $100$ times slower with respect to
sequential version due to synchronisation overheads (i.e. cache coherence,
cache misses, etc.).

All in all, intra-core SIMD parallelism appears  the only viable way
to this kind of parallelisation. Observe however that if it might
require,  for this class of algorithms, a  coding effort that easily
overcomes the potential benefits.

\begin{figure}
\begin{center}
\textsf{\small
\begin{tabular}{rrrcc}
\toprule
\# of species& Sequential (S)& SIMD (S)& Speedup&Ideal speedup\\
\midrule
2&       5.021& 5.071&0.99&4\\
4&      19.076& 18.887&1.01&4\\
8&       70.743& 70.043&1.01&4\\
16&      284.276& 278.701&1.02&4\\
32&      1121.231& 1099.245&1.02&4\\
\bottomrule
\end{tabular}
}
\end{center}
\caption{Execution time (S) and speedup of the SIMD CWC simulator against
  the sequential version on the $n$-species
  \emph{Lotka-Volterra}.\label{fig:simd:speedup}} 
\end{figure}

\subsection{Speeding up independent simulation instances}
\label{sec:implmany}
\begin{figure}
\begin{center}
\includegraphics[width=0.9\linewidth]{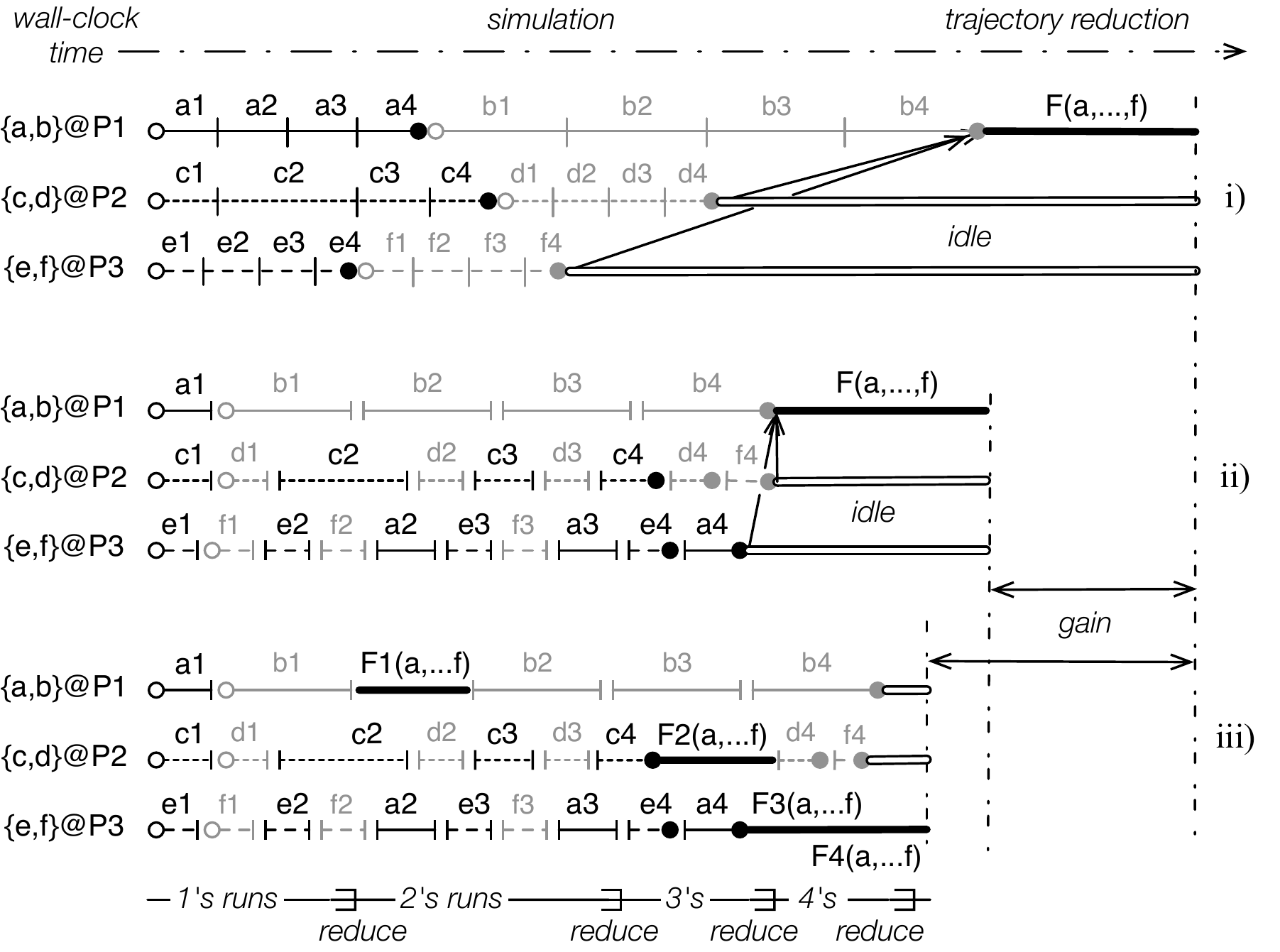}
\caption{Three alternative parallelisation schemas exemplified on  6
  simulation instances and 3 processors. \emph{i)} Round-robin execution of
  simulations followed by a reduction phase. \emph{ii)} Auto-balancing
  schema with time-slicing at constant simulation time (variable
  wall-clock time)  followed by a reduction phase. \emph{iii)} Previous
  schema with on-line pipelined reduction.\label{fig:parschema}}
\end{center}
\end{figure}

Starting from the CWC sequential simulator code sketched in
Fig.~\ref{fig:simd}, we here advocate a parallelisation schema
supporting the parallel execution of many self-balancing simulation instances on
multi-core. Its design aims to address all the issues discussed in
Sec.~\ref{sec:manysims}: it is realised by means of the \ff framework
(see Sec.~\ref{sec:fastflow}) that natively supports high-level parallel
programming patterns working on data streams and it exhibits an efficient
lock-free run-time support that can be integrated with SIMD code. It
therefore makes it possible the easy porting of the sequential CWC
code on multi-core for the execution of multiple
simulation instances (either replicas  or the parameter sweeping of a
simulation), and the on-line synthesis of their trajectories, which
can be made according to one or more associative reduction functions,
e.g. average, variance, confidence.  

The schema supports three main behaviours, which are exemplified in
Fig.~\ref{fig:parschema}:\\

\noindent
\emph{i)} The different simulation instances (called \helv{a,b,c,d,e,f}) are
dispatched for the execution on different workers threads of a \ff farm, which run
on different cores; a worker sequentially runs all the simulations it
received. The dispatching of instances to workers could be either
performed before the execution according to some static policy
(e.g. Round-Robin) or via an on-line scheduling policy 
(e.g. on-demand). Workers stream out the trajectories, which are
sampled along fixed steps along simulation time. Streams are buffered in the farm
collector and then reduced in a single stream according to one or more
functions (e.g. \helv{F}). Observe that the constant sampling assumption
simplify the reduction process even if it is not strictly required
since data could be on-line re-aligned during the buffering
\cite{stochkit-ff:tr-10-12}.  Also notice that since simulation time
advances according to a random variable, different instances advance
at different wall-clock time rates. The phenomenon is highlighted in
Fig.~\ref{fig:parschema} \emph{i)} splitting each instance in four
equal fractions of the simulation time (e.g. $\langle$\helv{a1, a2, a3,
  a4}{}$\rangle$, $\langle$\helv{b1, b2, b3, b4}{}$\rangle$, which
exhibit different wall-clock time to be computed (segment
length). This may induce even a significant load unbalance that could
be only partially addressed using on-line scheduling policies.\\

\noindent
\emph{ii)} A possible solution to improve load balancing of the schema
consists in coupling the on-line scheduling policy with the reduction
of execution time-slice that is subject to the scheduling policy. At
this end, each simulation instance can be represented as an object
that incorporate its current progress and provide the scheduler with
the possibility of stopping and restarting an instance. In this way,
as it happens in a  time-sharing operating system, (fixed or variable
length) slices of an instance can be scheduled on different workers
provided slices of the same instances are sequenced (possibly on
different workers). Thanks to cache-coherent shared memory the
scheduling can be efficiently realised via pointer management. The
idea is exemplified in  Fig.~\ref{fig:parschema} \emph{ii)}. Also,
scheduling and dispatching to workers can be equipped with predictive
heuristics based on instance history in order to characterise the
relative speed of the simulation instances.\\

\noindent
\emph{iii)} The previous schema can be further improved by pipelining the
reduction phase that is performed on-line. Since instance time-slicing
can make all the instances to 
progress, a running window of all the trajectories can be reduced
while they are still being produced. The reduction process, which is
logically made within a separate thread (i.e. the farm
collector), can be either run on an additional processor or
interleaved with the execution of simulation instances (see 
Fig.~\ref{fig:parschema} \emph{iii)}. The solution also significantly reduces
the amount of data to be kept in memory because: 1) thanks
to interleaving all the trajectories advances almost aligned with
respect to simulation time; 2)  the already reduced parts of the
trajectories can be deleted from main memory (and stored in secondary
memory if needed).\\

\begin{figure}
\begin{center}
\includegraphics[width=0.4\linewidth]{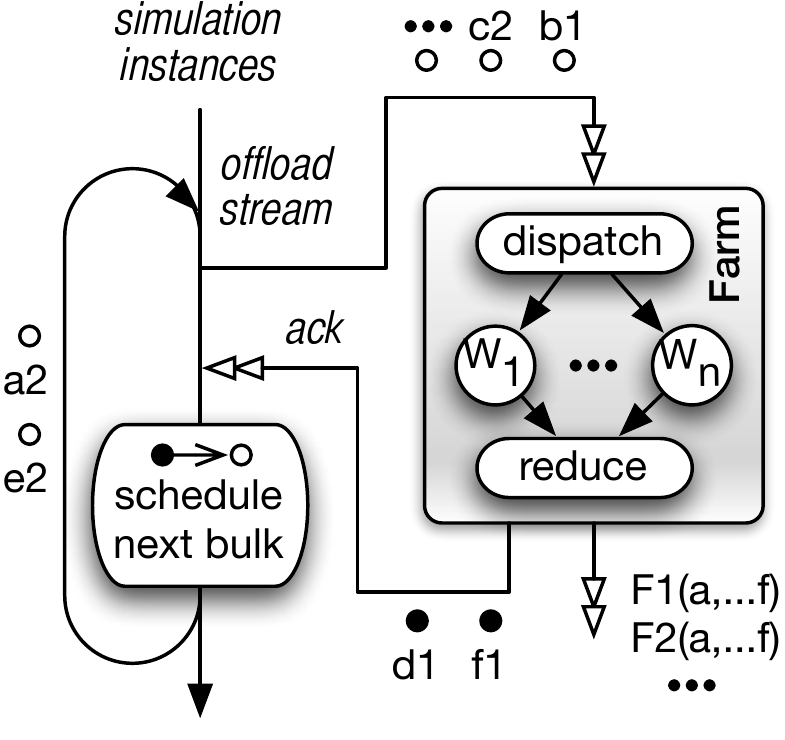}
\caption{Architecture of the \ff-based CWC parallel simulator.\label{fig:simarch}}
\end{center}
\end{figure}

 The three schemas can be effectively implemented using \ff as sketched in
 Fig.~\ref{fig:simarch}. In particular, the \ff \emph{farm accelerator}
 feature \cite{fastflow:web} 
 fits well the previous design since it makes possible to
 offload\footnote{So-called \emph{self-offload} since the accelerator use
   the same hardware device of the main thread, i.e. multi-core CPU,
   see Fig.~\ref{fig:ff:architecture}  \cite{fastflow:web}.} a
 stream of object pointers onto a farm of workers, each of them running a CWC
 simulator, and to implement user-defined dispatching and reduction
 functions via standard Object Oriented subclassing. 
As discussed in
 Sec.~\ref{sec:fastflow}, \ff natively provides the programmer with
 streams, a configurable farm pattern, and an efficient run-time
 support based on lock-free synchronisations. All these features
 effectively made it possible to port the CWC sequential simulator to
 multi-core with moderate effort (few days for our team of
 students). In addition, the complexity of the achieved solution can
 be gracefully improved by successive refinements in order to test
 different scheduling policies or variants to the basic schema. In
 this regard the accelerator feature represents a key issue since it
 enables the programmer to make very local changes to the original
 code that in first approximation consists in changing a method call
 into the offload of the same method.
 
%According to the schema iii, we
%provided it with basic scheduling concepts and online result-streams
%reduction.

\begin{figure*}
\begin{center}
\includegraphics[width=0.75\linewidth]{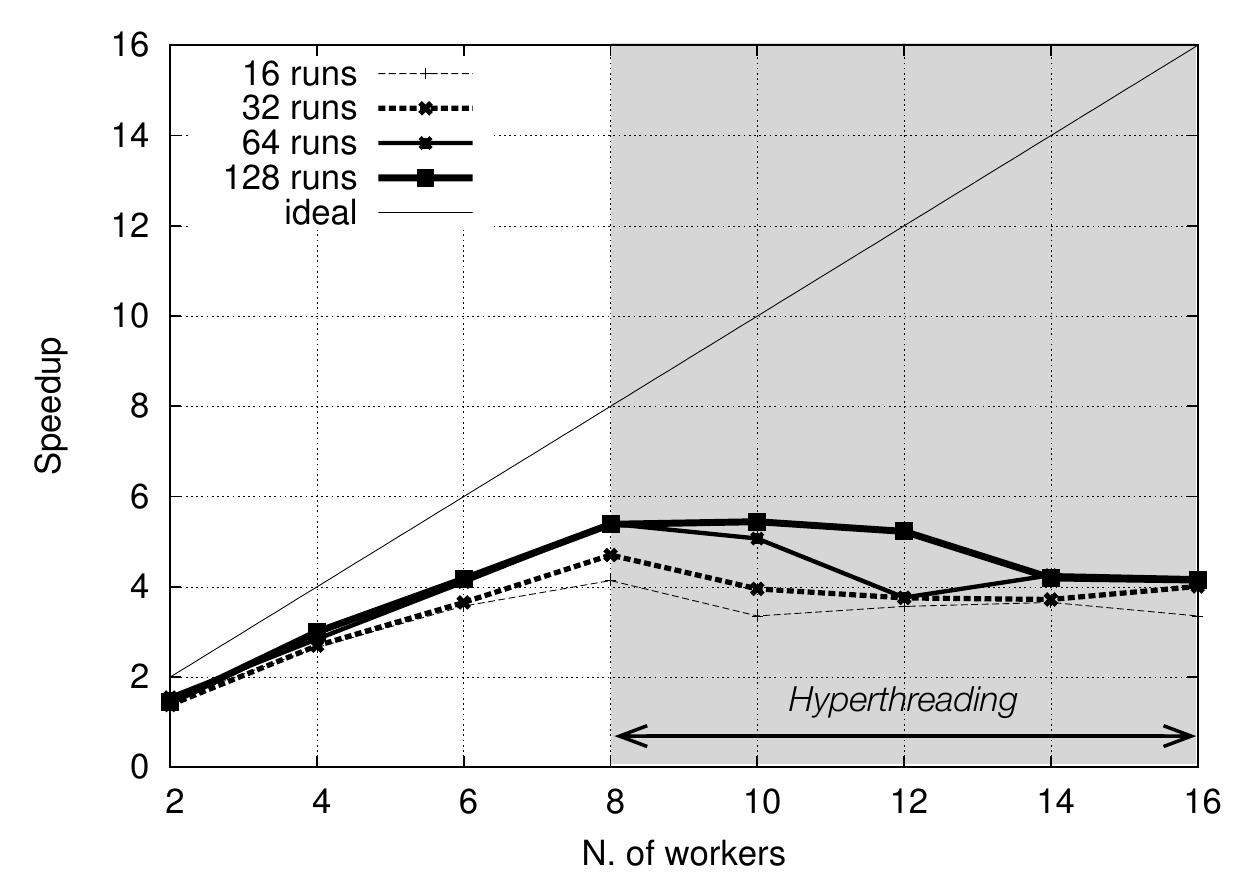}
\hfill
\includegraphics[width=0.75\linewidth]{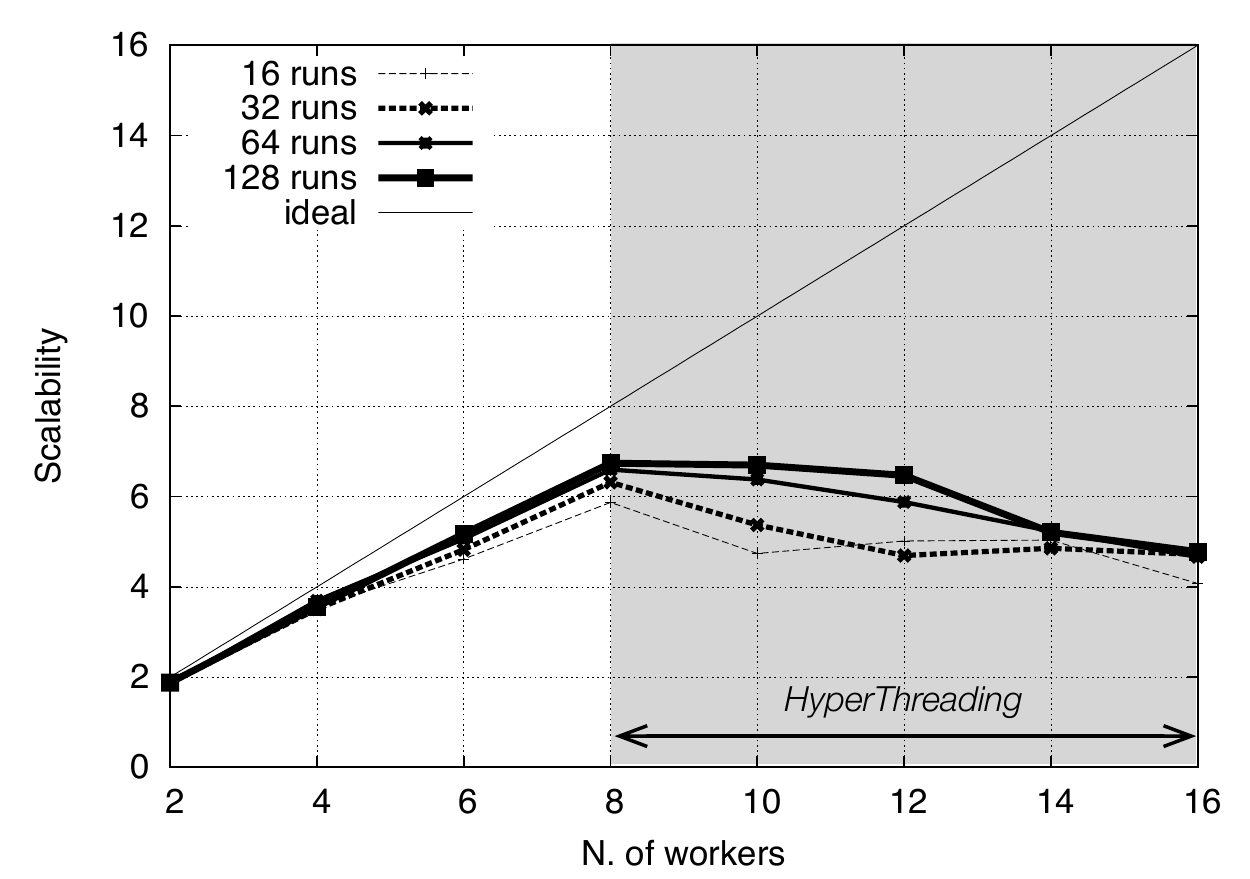}
\caption{Speedup ($T_{par(n)}$ vs $T_{seq}$) and scalability
  ($T_{par(n)}$ vs $T_{par(1)}$) of the parallel CWC simulator, see Sec.~\ref{sec:implmany}  \emph{iii)}.\label{fig:speedup}} 
\end{center}
\end{figure*}

Figure~\ref{fig:speedup} reports the achieved
speedup and scalability for schema \emph{iii)}, evaluated on multiple
instances of simulation over the 
Lotka-Volterra model. In the speedup plot, the measures for the
parallel version includes the time spent for 
computing reductions (means, variance, and confidence), whereas this
cost is not measured in the sequential version as it is managed as
post-processing phase; nevertheless, the achieved speedup is
reasonably good.  The quality of parallelisation work is further
confirmed by the scalability plot, where the parallel version is
compared with a sequential code that  computes the same
reductions (means, variance, and confidence). 
%in which the time spe times of
%the parallel version ($T_{par(n)}$) are compared with a sequential
%version that also computes the same reductions ($T_{par(1)}$). 
In this
case the achieved scalability is close to the ideal one. 
Observe that, as typical in memory-bound
 CPU-intensive workloads, HyperThreading does not bring any additional
 benefit.

\section{Related Works}

The parallelisation of stochastic simulators has
been extensively studied in the last two decades. Many of these
efforts focus on distributed architecture and specific
simulators. Our work differs from these efforts in three main aspects:
1) it mainly address multicore-specific parallelisation issues; 2)
it advocates a general parallelisation schema rather than a specific
simulator, 3) it specifically address the on-line reduction of
simulation trajectories, thus it is designed to manage large streams
of data. To the best of our knowledge, many related works covers some
of these aspects, but very few of them (if any) address all three
aspects. Among related works, some are worth to be explicitly  mentioned. 

The Swarm algorithm  \cite{RS01}, which is 
well suited for biochemical pathway optimisation has been used in a
distributed environment, e.g., in Grid Cellware~\cite{Detal05}, a
grid-based modelling and simulation tool 
for the analysis of biological pathways that offers an integrated
environment for several mathematical representations ranging from
stochastic to deterministic algorithms. 

Parameter Sweep Applications (PSAs) exploit that aim  must involve
making the problem very time consuming. However, since the
instances of a PSA are independent, the distributed computing paradigm
to to sample a large space of independent instances.  
In \cite{MCMPMM09}, a grid-based version of a multi-volume stochastic
simulator is presented.  

%In \cite{BBCDFS09}, the authors present a prototype tool for parallel (distributed) analysis of multi-affine systems. 
%Abstraction refinement techniques are also implemented.

DiVinE \cite{BBCS05} is a general distributed verification environment meant to support the development of distributed enumerative model checking algorithms.
%A tool for parallel shared memory enumerative
%LTL model checking and reachability analysis is provided, based on
%distributed memory algorithms implemented for multi-core
%and multi-cpu environments using shared memory.
In \cite{BBC+08}, features about probabilistic analysis have been
added to DiVinE. The model checker has been massively used for the
analysis of biological systems, see, e.g., \cite{BBS10}

StochKit \cite{stochkit:web}  is an extensible
stochastic simulation framework developed in the C++ language. It
aims at making stochastic simulation accessible to biologists, while
remaining open to extension via new stochastic and  
multi-scale algorithms. It implements the Gillespie
algorithm, and other methods. It is a programming
framework and in its second version it targets multi-core platforms,
it is therefore similar to our work. It does not implement any kind
of SIMD parallelism nor on-line trajectory reduction (that is
performed as post-processing). 
%StochKit-FF \cite{stochkit-ff:tr-10-12}, 

\section{Concluding remarks}

Starting from the Calculus of Wrapped Compartments  we have discussed the
main parallelisation issues for its simulator, and in general for the stochastic 
simulation of biological systems,  on commodity multi-core
platforms. In particular, we distinguished two different approaches to
parallelisation, i.e. the parallelisation of the single simulation
instance and many simulation instances.  For each class we have
defined a number of design guidelines, which we believe, may support
the easy and efficient porting of this class of algorithms on
multi-cores. These guidelines include both the programming language
abstractions (streams and high-level programming patterns), the
run-time mechanisms (intra-core SIMD parallelism, lock-free
cache-friendly inter-core synchronisations here provided by the \ff
framework), and basic simulator 
architectural schema (simulation ``objectification'', interleaved
execution and pipelined reduction), which can be gracefully optimised
with limited effort to experiment different parallel execution behaviours. 

The presented guidelines have been used to develop a multicore-aware
porting of the CWC simulator, which have been experimented over classic
simulation problems. The experimental evidence obtained in the design and
utilisation of the parallel simulator are convincing both in term of
the achieved performance and the moderate porting effort for the
parallelisation of multiple instances whereas it appears disappointing
for the parallelisation of the single instance.

Both the \ff framework and the CWC simulator are open source software
under LGPL licence \cite{fastflow:web,cwc:web}.

\section{Acknowledgements}
\thanks{This work has been partially
  supported by a grant of the HPC-Europa 2 {\em Transnational Access}
  programme (grant n. 228398) and by the {\em
    BioBITs} (``Developing White and Green Biotechnologies by Converging Platforms from Biology and Information Technology towards Metagenomic''). 

The final, revised version of this paper will appear in Proc. of the 19th
Euromicro Intl. Conf. on Parallel, Distributed and
Network-Based Computing (PDP), Ayia Napa, Cyprus,
Feb. 2011. IEEE. \url{http://ieeexplore.ieee.org}

%\bibliographystyle{abbrv}
%\bibliography{UniPisaGroup,multicore,fmb,ac}

\begin{thebibliography}{10}

\bibitem{stochkit-ff:tr-10-12}
M.~Aldinucci, A.~Bracciali, P.~Li{\`o}, A.~Sorathiya, and M.~Torquati.
\newblock {StochKit-FF}: Efficient systems biology on multicore architectures.
\newblock Technical Report TR-10-12, Universit{\`a} di Pisa, Dipartimento di
  Informatica, Italy, 2010.

\bibitem{lithium:sem:CLSS}
M.~Aldinucci and M.~Danelutto.
\newblock Skeleton based parallel programming: functional and parallel semantic
  in a single shot.
\newblock {\em Computer Languages, Systems and Structures}, 33(3-4):179--192,
  Oct. 2007.

\bibitem{fastflow:pdp:10}
M.~Aldinucci, M.~Meneghin, and M.~Torquati.
\newblock Efficient {Smith-Waterman} on multi-core with fastflow.
\newblock In {\em Proc. of Intl. Euromicro PDP 2010: Parallel Distributed and
  network-based Processing}, pages 195--199, Pisa, Italy, Feb. 2010.

\bibitem{ABI01}
R.~Alur, C.~Belta, and F.~Ivancic.
\newblock Hybrid modeling and simulation of biomolecular networks.
\newblock In {\em HSCC}, volume 2034 of {\em LNCS}, pages 19--32. Springer,
  2001.

\bibitem{amdahl:law:1967}
G.~M. Amdahl.
\newblock Validity of the single processor approach to achieving large scale
  computing capabilities.
\newblock In {\em AFIPS '67 (Spring): Proc. of the April 18-20, 1967, spring
  joint computer conference}, pages 483--485, New York, NY, USA, 1967. ACM.

\bibitem{BBS10}
J.~Barnat, L.~Brim, and D.~Safr{\'a}nek.
\newblock High-performance analysis of biological systems dynamics with the
  divine model checker.
\newblock {\em Briefings in Bioinformatics}, 11(3):301--312, 2010.

\bibitem{BBC+08}
J.~Barnat, L.~Brim, I.~\v{C}ern\'{a}, M.~\v{C}e\v{s}ka, and J.~T\r{u}mov\'{a}.
\newblock {ProbDiVinE-MC: Multi-core LTL Model Checker for Probabilistic
  Systems}.
\newblock In {\em QEST '08: Proceedings of the 2008 Fifth International
  Conference on Quantitative Evaluation of Systems}, pages 77--78, Washington,
  DC, USA, 2008. IEEE Computer Society.

\bibitem{BBCS05}
J.~Barnat, L.~Brim, I.~\v{C}ern\'{a}, and P.~\v{S}ime\v{c}ek.
\newblock {D}i{V}in{E} -- {T}he {D}istributed {V}erification {E}nvironment.
\newblock In M.~Leucker and J.~van~de Pol, editors, {\em Proceedings of 4th
  International Workshop on Parallel and Distributed Methods in verifiCation},
  pages 89--94, July 2005.

\bibitem{bernstein:66}
A.~J. Bernstein.
\newblock Program analysis for parallel processing.
\newblock {\em IEEE Trans. on Electronic Computers}, EC-15(5):757--762, 1966.

\bibitem{Car05}
L.~Cardelli.
\newblock Brane calculi.
\newblock In {\em CMSB}, volume 3082 of {\em LNCS}, pages 257--278. Springer,
  2005.

\bibitem{cole-th}
M.~Cole.
\newblock {\em Algorithmic Skeletons: Structured Management of Parallel
  Computations}.
\newblock Research Monographs in Parallel and Distributed Computing. Pitman,
  1989.

\bibitem{preQAPL2010}
M.~Coppo, F.~Damiani, M.~Drocco, E.~Grassi, and A.~Troina.
\newblock {S}tochastic {C}alculus of {W}rapped {C}ompatnents.
\newblock In {\em QAPL'10}, volume~28, pages 82--98. EPTCS, 2010.

\bibitem{mapreduce:google:04}
J.~Dean and S.~Ghemawat.
\newblock {MapReduce}: Simplified data processing on large clusters.
\newblock In {\em Usenix OSDI '04}, pages 137--150, Dec. 2004.

\bibitem{DPR08}
L.~Dematt{\'e}, C.~Priami, and A.~Romanel.
\newblock The beta workbench: a computational tool to study the dynamics of
  biological systems.
\newblock {\em Briefings in Bioinformatics}, 9(5):437--449, 2008.

\bibitem{Detal05}
P.~K. Dhar and et~al.
\newblock Grid cellware: the first grid-enabled tool for modelling and
  simulating cellular processes.
\newblock {\em Bioinformatics}, 7:1284--1287, 2005.

\bibitem{DBLP:conf/iccms/DittamoC09}
C.~Dittamo and D.~Cangelosi.
\newblock Optimized parallel implementation of {Gillespie's} first reaction
  method on graphics processing units.
\newblock In {\em International Conference on Computer Modeling and Simulation,
  (ICCMS)}, pages 156--161, Macau, China, Feb. 2009. IEEE.

\bibitem{balbo:sim:1998}
A.~Ferscha.
\newblock {\em Performance Models for Discrete Event Systems with
  Synchronisations: Formalisms and Analysis Techniques}, volume~2 of {\em MATCH
  Advanced Schools}, chapter VII -- Simulation.
\newblock Jaca, Spain, Sept. 1998.

\bibitem{G77}
D.~Gillespie.
\newblock Exact stochastic simulation of coupled chemical reactions.
\newblock {\em J. Phys. Chem.}, 81:2340--2361, 1977.

\bibitem{Herlihy91}
M.~Herlihy.
\newblock Wait-free synchronization.
\newblock {\em ACM Trans. Program. Lang. Syst.}, 13(1):124--149, 1991.

\bibitem{intel:sse:intrics}
Intel.
\newblock {\em Intel \textregistered\ C++ Intrinsics Reference}, 2010.

\bibitem{intel:skeletons:tbb}
Intel Corp.
\newblock {\em Threading Building Blocks}, 2009.
\newblock \url{http://www.threadingbuildingblocks.org/}.

\bibitem{KMT08}
J.~Krivine, R.~Milner, and A.~Troina.
\newblock Stochastic bigraphs.
\newblock {\em ENTCS}, 218:73--96, 2008.

\bibitem{MCMPMM09}
E.~Mosca, P.~Cazzaniga, I.~Merelli, D.~Pescini, G.~Mauri, and L.~Milanesi.
\newblock Stochastic simulations on a grid framework for parameter sweep
  applications in biological models.
\newblock {\em High Performance Computational Systems Biology, International
  Workshop on}, pages 33--42, 2009.

\bibitem{stochkit:web}
L.~Petzold.
\newblock {\em {StochKit}: stochastic simulation kit web page}, 2009.
\newblock \url{ http://www.engineering.ucsb.edu/~cse/StochKit/index.html}.

\bibitem{PRSS01}
C.~Priami, A.~Regev, E.~Y. Shapiro, and W.~Silverman.
\newblock Application of a stochastic name-passing calculus to representation
  and simulation of molecular processes.
\newblock {\em Inf. Process. Lett.}, 80(1):25--31, 2001.

\bibitem{P02}
G.~P\v{a}un.
\newblock {\em Membrane computing. An introduction}.
\newblock Springer, 2002.

\bibitem{RS01}
T.~Ray and P.~Saini.
\newblock Engineering design optimization using a swarm with an intelligent
  information sharing among individuals.
\newblock {\em Eng. Opt.}, 33:735--748, 2001.

\bibitem{fastflow:web}
Sourceforge website.
\newblock {\em FastFlow project}, 2009.
\newblock \url{http://mc-fastflow.sourceforge.net/}.

\bibitem{cwc:web}
Sourceforge website.
\newblock {\em CWC Simulator project}, 2010.
\newblock \url{http://sourceforge.net/projects/cwcsimulator/}.

\end{thebibliography}

\end{document}